\documentstyle[preprint,aps]{revtex}
\draft
\begin{document}
\bibliographystyle{prsty}
% \bibliographystyle{abbrv}
%%%%%%%%%%%%%%%%%%%%%%%%%%%%%%%%%%%%%%%%%%%%%%%%%%%%%%%%%%%%%%%%%%%%%%%%%%%
\preprint{RUB-TPII-18/95}
\title{Strangeness in the Scalar Form Factor of the Nucleon}

\author{
Hyun-Chul Kim
\footnote{E-mail address:kim@hadron.tp2.ruhr-uni-bochum.de},
Andree Blotz
\footnote{Present address: Department of Physics, State University
of New York, Stony Brook, 11794, U.S.A.},
C. Schneider
\footnote{E-mail address:carstens@elektron.tp2.ruhr-uni-bochum.de},
and Klaus Goeke
\footnote{E-mail address:goeke@hadron.tp2.ruhr-uni-bochum.de}}
\address{
Institute for  Theoretical  Physics  II, \\  P.O. Box 102148,
Ruhr-University Bochum, \\
 D-44780 Bochum, Germany  \\
       }
\date{July, 1995}
\maketitle
\begin{abstract}
The scalar form factor of the nucleon and related physical quantities
are investigated in the framework of the semibosonized
SU(3) Nambu-Jona-Lasinio soliton model.
We take into account the rotational
$1/N_c$ corrections and linear $m_s$ corrections.
The strangeness content of the nucleon in the scalar form factor
is discussed in detail.  In particular, it is found that
the $m_s$ corrections play an essential role of reducing the
$\langle N |  \bar{s} s | N \rangle$
arising from the leading order and rotational
$1/N_c$ contributions.  We obtain the
$\sigma_{\pi N} (0)=40.80\;\mbox{MeV}$,
$\Delta \sigma = \sigma_{\pi N} (2m^{2}_{\pi})-\sigma_{\pi N}
(0) = 18.18\;\mbox{MeV}$ and
$\langle r^2\rangle^{S}_{N} = 1.50\;\mbox{fm}^2$.
 The results are in a remarkable agreement with empirical data
analyzed by Gasser, Leutwyler, and Sainio~\cite{gls}.
\end{abstract}

\section{Introduction}
Since Cheng ~\cite{cheng} showed that there is a
factor-of-two discrepancy between
the empirical data for the pion-nucleon sigma term ($\Sigma_{\pi N}$) and
the naive estimates of the $\sigma$-term  from the mass spectrum,
there have been a great deal of discussions and
disputes about the $\Sigma_{\pi N}$ and $\sigma$ term
(see Ref.~\cite{jk,gls} and references therein).
Donoghue and Nappi~\cite{dn} suggested that the discrepancy
is due to the presence of strange quarks in the nucleon, {\em i.e.}
$\langle N| \bar{s}s |N\rangle\neq 0$ and showed that
$\langle N| \bar{s}s |N\rangle$ contributes almost $30\%$ to the
quark condensate in the nucleon, making use of the Skyrme model
and bag model.  At the first thought, it seems to be reasonable,
since Cheng used the Zweig rule, {\em i.e.} neglected
$\langle N| \bar{s}s |N\rangle$.
However, one serious question arises: a large fraction of the
nucleon mass then stems from strange quarks if one follows
Ref.~\cite{dn}, which contradicts the quark model.
Another assumption was that the ratio $m_s/\bar{m}$ is off
by a factor of two, which means that the first order
perturbation theory collapses.  However, this kind of
suggestion would lead to a breakdown of the
Gell-Mann-Okubo mass formula which predict the masses of
hadrons in a few percent.

Motivated by these contradictions,
Gasser, Leutwyler and Sainio~\cite{gls} recently
reanalysed the $\sigma$ term prudently, taking advantage
of newly accumulated and better $\pi N$ scattering data
and considering the strong $t$-dependence of the scalar
form factor $\sigma(t)$
($\sigma(2m^{2}_{\pi})-\sigma (0)\simeq 15 \;\mbox{MeV}$).
The results of Ref.~\cite{gls} were $\sigma = 45 \pm 8 \mbox{MeV}$
and $\Sigma \simeq 60 \mbox{MeV}$.
The $y=2\langle N| \bar{s}s |N\rangle/\langle N|
\bar{u}u + \bar{d}d |N\rangle$, a share
of $\langle N| \bar{s}s |N\rangle$ in the $\sigma$ term,
was about 0.2, so that the corresponding contribution of the
term $\langle N| \bar{s}s |N\rangle$ to the nucleon mass was
about $130\;\mbox{MeV}$.

In the meanwhile, the efforts to understand the $\sigma$ term puzzle
theoretically have continued~\cite{bft,bass,sw}.  However, the bone of
contention still lies in the role of strange quarks, more specifically
the contribution of the $\langle N| \bar{s}s |N\rangle$ to the
$\sigma$ term.  Recently, several works insist that
there is no need to introduce a portion of strange quarks to
explain the $\sigma$ term discrepancy.
Bass~\cite{bass} proposed that based on the
Gribov confinement the value of the $\sigma$
term can be explained without need to invoke large strangeness
content of the nucleon.  Ball, Forte and Tigg~\cite{bft} also
suggested that with the correct understanding of the baryon matrix
element the $\sigma$ term (identified with
$\sigma_8 =\bar{m} \langle N| \bar{u}u +\bar{d}d -2\bar{s}s
| N \rangle$ ) can be reproduced without
violating the Zweig rule.  Hence, following these arguments,
strange quarks do not contribute to the nucleon mass.
Though it should be small, it is still important to consider
the contribution of strange quarks to the $\sigma$ term,
in line with recent experiments indicating the fact
that strange quarks might play an important role of explaining
the properties of the nucleon~\cite{emc,bnl}.

It is the object of the present work to study the strangeness
contribution to the $\sigma$ term in the framework of the
semi-bosonized SU(3) Nambu-Jona-Lasinio soliton model (often called
as the chiral quark soliton model).
In our model, the nucleon is understood explicitly as
$N_c$ valence quarks coupled to the polarized Dirac sea
bound by a non-trivial chiral mean field configuration.
The proper quantum numbers of the nucleon can be acquired by
the semiclassical quantization~\cite{dpp,anw} performed
via integrating over the zero-mode fluctuations of the pion field
around the saddle point.  It allows the nucleon to carry proper
quantum numbers such as spins and isospins.
The SU(3) NJL soliton model has a merit in that it interpolates between
the nonrelativistic naive quark model and the Skyrme model.
It enables us to study the interplay between these two
different models~\cite{PraBlGo}.
The model is quite successful in describing the
static properties of the baryons and their form factors
\cite{betal,bpg,kbpg}.

The outline of the paper is as follows:
In the next section, we sketch the basic formalism for
the scalar form factor in SU(3) NJL soliton the model.
In section 3, we present the numerical results and
discuss about them.
In section 4, we summarize the present work and remark the conclusion.

\section{Formalism}
The scalar form factor $\sigma (t)$ is defined as a condensate of
$u$ and $d$ quarks in the nucleon:
\begin{equation}
\sigma (t) \;=\; \bar{m}
\langle N(p') |\bar{u}u + \bar{d}{d} | N(p) \rangle
\label{Eq:sigma1}
\end{equation}
with $\bar{m}=(m_u+m_d)/2\simeq6\;\mbox{MeV}$.
The $t$ denotes the square of the momentum transfer.
Our model is characterized by a
low--momenta QCD partition function in Euclidean space
given by the functional integral over pseudoscalar meson
and quark fields:
\begin{equation}
{\cal Z} \; = \; \int {\cal D} \Psi {\cal D} \Psi^{\dagger}
{\cal D} \pi^A \exp{\left(-\int d^4x\Psi^\dagger iD\Psi\right)}
\label{Eq:func}
\end{equation}
where
\begin{equation}
iD \;=\; \beta(-i\rlap{/}{\partial}\;+\;
MU^{\gamma_5}\;+\;\hat{m})
,\;\;U^{\gamma_5}=e^{i\pi^a\lambda^a\gamma_5}.
\end{equation}
$\lambda^a$ are SU(3) Gell-Mann matrices normalized as
$\mbox{Tr}{\lambda^a\lambda^b} = 2\delta^{ab}$.  The $\hat{m}$
denotes the current quark mass matrix for which
we take the form $diag(m_u,m_d,m_s)$, where
$m_u, m_d$ and $m_s$ are the corresponding current quark masses
of the {\em up}, {\em down} and {\em strange} quark, respectively.
Here, we assume that isospin symmetry is not broken, {\em i.e.}
$m_u=m_d=\bar{m}$.
The $M$ stands for the momentum--dependent dynamical mass arising
from the spontaneous chiral symmetry breaking.
The momentum--dependence of the $M$ introduces the ultra--violet
cut--off.  However, we shall regard it as a constant for simplicity.
Instead, we employ a simple proper--time regularization.
The differential operator $iD$ is expressed in Euclidean space in terms of
the Euclidean time derivative $\partial_\tau$, the Dirac one-particle
Hamiltonian $H(U)$ and symmetry breaking part~\cite{sbg}:
\begin{equation}
iD\;=\;\partial_\tau \;+\; H(U) \;+\; h_{sb}
\end{equation}
with
\begin{equation}
H(U)\;=\;\frac{\vec{\alpha}\cdot \nabla}{i} \;+\; \beta M_u U
\;+\; \beta \bar{m}{\bf 1},\;\;\; h_{sb} \;=\;\beta \mu_0 {\bf 1}
\;+\;\beta \mu_8 \lambda_8.
\end{equation}
Here, we have made the famous embedding Ansatz for the pseudoscalar
fields $U^{\gamma^5}$ and
$U$ is expressed by
\begin{equation}
U\;=\; \left(\begin{array}{cc} U_{0} & 0 \\
0 & 1
\end{array} \right).
\end{equation}
The $U_{0}$ expresses the SU(2) chiral background field
$U_0=\exp{i[\vec{n}\cdot\vec{\tau} P(r)]}$ with the hedgehog Ansatz.
$P(r)$ denotes the profile function with proper boundary conditions.
$\mu_0$ and $\mu_8$ are defined by
$\mu_0 = (M_s - M_u) / 3$ and $\mu_8 = -(M_s - M_u) / \sqrt{3}$.
$M_s$ and $M_u$ are constituent quark masses of the $s$ and $u$ quarks
respectively.  The $M_u$ is used as an input parameter, while
the $M_s$ is determined by the gap equation~\cite{sbg}.
The current strange quark mass $m_s$ is also settled in the same way.
We treat the explicit symmetry breaking term $h_{sb}$  perturbatively.
 The hadronic matrix elements of the $\pi\mbox{N}\; \sigma$--term
is related to the correlation function
\begin{equation}
\sigma(t)\;\smash{\mathop{\sim}\limits_{T\rightarrow \infty}}\;
\langle 0 | J_N (\vec{x}, \frac{T}{2})\hat{\sigma}
J^{\dagger}_{N} (\vec{y}, -\frac{T}{2}) | 0 \rangle
\label{Eq:sigma}
\end{equation}
at large Euclidean time $T$.  $\hat{\sigma}$ is the quark operator
for the $\sigma$ term, defined by
$\hat{\sigma}=\bar{m}(\bar{u}u+\bar{d}d)$.
$J_N$ is the nucleon current
constructed from $N_c$ quark fields~\cite{dpp}
\begin{equation}
J_N(x)\;=\; \frac{1}{N_c !} \epsilon_{i_1 \cdots i_{N_c}}
\Gamma^{\alpha_1 \cdots
\alpha_{N_c}}_{JJ_3TT_3Y}\psi_{\alpha_1i_1}(x)
\cdots \psi_{\alpha_{N_c}i_{N_c}}(x).
\end{equation}
$\alpha_1 \cdots\alpha_{N_c}$ denote spin--flavor indices, while
$i_1 \cdots i_{N_c}$ designate color indices.  The matrices
$\Gamma^{\alpha_1 \cdots\alpha_{N_c}}_{JJ_3TT_3Y}$ are taken to endow
the corresponding current with the quantum numbers $JJ_3TT_3Y$.
 The $J^{\dagger}_{B}$ plays the role of creating the baryon state.

The integral over the quark fields are trivial.
The integral over the pseudo-Goldstone boson fields can be performed
by the saddle point method in the large $N_c$ limit.
In order to find the quantum $1/N_c$ corrections,
it is important to take into account the small oscillations of the
pseudo-Goldstone bosons around the saddle point and the zero modes.
The zero modes are taken into account by the
soliton expressed by $\tilde{U} (\vec{x},x_4)
=A(x_4)U(\vec{x} - \vec{Z})A^{\dagger} (x_4)$
with an SU(3) unitary matrix $A(t)$.
 Hence, the collective action $S_{eff}$ becomes
\begin{eqnarray}
\tilde{S}_{eff} & = & -N_c {\rm Sp} \log{(iD)} \nonumber \\
& = & -N_c \mbox{Sp}
\log{\left [ \partial_\tau \;+\; H(\tilde{U})
\;+\; A^{\dagger} (x_4) \dot{A}(x_4)
\;-\;  i \beta \dot{\vec{Z}} \cdot \nabla \right. }
\nonumber \\
& & \; +\;  \left. A^{\dagger}(x_4) h_{sb} A(x_4)
\;-\; \xi (y) \beta A^{\dagger}(x_4) \frac{1}{\sqrt{3}}
(\sqrt{2}\lambda_0 + \lambda_8) A(x_4) \right]
\end{eqnarray}
with the angular velocity $A^\dagger (x_4) \dot{A} (x_4) = i\Omega_E
=i\Omega^{a}_{E} \lambda^a /2$.  $\mbox{Sp}$ denotes the
functional trace.
The $\xi$ stands for
the external scalar field, with regard to which
we make a functional derivative so as to obtain the sigma form factor:
\begin{eqnarray}
\sigma(t) &=& -N_c
\frac{\delta }{\delta \xi(z)} \mbox{Sp}
\log{\left \{ \partial_\tau \;+\; H(\tilde{U})
\;+\; A^{\dagger} (x_4) \dot{A}(x_4)
\;-\;  i \beta \dot{\vec{Z}} \cdot \nabla \right. }
\nonumber \\
& & \; +\; \left. A^{\dagger}(x_4) h_{sb} A(x_4)
\;-\; \xi (y) \beta A^{\dagger}(x_4) \frac{1}{\sqrt{3}}
(\sqrt{2}\lambda_0 + \lambda_8) A(x_4) \right \}
\end{eqnarray}
It is known that there is the dependence of
the $\sigma$ term on the regularization scheme~\cite{admg}.
However, we want to stress the fact that we have employed
the proper-time regularization and have evaluated
possible physical observables such as mass splittings,
magnetic moments, axial constants and electromagnetic form factors
within the same scheme and same values of input parameters
\footnote{In fact,
we have only one free parameter, {\em i.e.} the constituent
up-quark (down-quark) mass.  However, it is more or less fixed
to around $420\;\mbox{MeV}$ by
the mass splitting~\cite{betal}.  }.
Hence, we stick to
the proper-time regularization for the $\sigma$ term
and make use of the same input parameters without adjusting.
However, we shall not be here bothered by going through all the
tedious technical details arising from the regularization
(see Ref.~\cite{kpbg} for details).

Having taken into account the rotational $1/N_c$ corrections and
linear $m_s$ corrections, we arrive at
\begin{eqnarray}
\sigma (t) & = & \Sigma_{SU(2)} (t) \langle 2 \;+\; D^{(8)}_{88}(A)
\rangle_N  \nonumber \\
& + & \frac{2\bar{m}}{\sqrt{3}I_1} {\cal K}_1 (t)
\langle  D^{(8)}_{8i}(A)R_i  \rangle_N
\; + \; \frac{2\bar{m}}{\sqrt{3} I_2} {\cal K}_2 (t)
\langle  D^{(8)}_{8p}(A)R_p  \rangle_N \nonumber \\
& - & \frac{4\bar{m}\mu_8}{\sqrt{3}}
\left[{\cal N}_1(t) - {\cal K}_1 (t) \frac{K_1}{I_1}\right]
\langle D^{(8)}_{8i}(A)D^{(8)}_{8i}(A)  \rangle_N
\nonumber \\
& - & \frac{4\bar{m}\mu_8}{\sqrt{3}}
\left[{\cal N}_2(t) - {\cal K}_2 (t) \frac{K_2}{I_2}\right]
\langle
D^{(8)}_{8p}(A)D^{(8)}_{8p}(A)  \rangle_N  \nonumber \\
&-& \frac{4\bar{m}\mu_8}{3\sqrt{3}}
{\cal N}_0 (t)
\langle D^{(8)}_{88}(A) (D^{(8)}_{88}(A) + 1) \rangle_N
\; - \; \frac{8\bar{m}\mu_0}{3}  {\cal N}_0 (t),
\label{Eq:sf}
\end{eqnarray}
where
\begin{eqnarray}
\Sigma_{SU(2)} (t) & = & N_c \int d^3x \;j_0 (Qr)
\left [\Psi^{\dagger}_{val} (x) \beta \Psi_{val} (x) \;-\; \sum_n
\frac{1}{2} \mbox{sign}(E_n) {\cal R} (E_n)
 \Psi^{\dagger}_{n} (x) \beta \Psi_n (x) \right], \nonumber \\
{\cal K}_1 (t) & = & \frac{N_c}{6} \sum_{n,m}
\int  d^3x j_0 (Qr) \int d^3y \left[
\frac{\Psi^{\dagger}_{n} (x) \vec{\tau} \Psi_{val} (x) \cdot
\Psi^{\dagger}_{val} (y) \beta \vec{\tau} \Psi_{n} (y)}
{E_n - E_{val}} \right .
\nonumber \\  & & \hspace{3cm} \;+\; \left . \frac{1}{2}
\Psi^{\dagger}_{n} (x) \vec{\tau} \Psi_{m} (x) \cdot
\Psi^{\dagger}_{m} (y) \beta \vec{\tau} \Psi_{n} (y)
{\cal R}_{\cal M} (E_n, E_m)
\right ],
\nonumber \\
{\cal K}_2 (t) & = & \frac{N_c}{6} \sum_{n, m^{0}}
\int d^3 x \;j_0 (Qr) \int d^3 y
\left [\frac{\Psi^{\dagger}_{m^{0}} (x) \Psi_{val} (x)
\Psi^{\dagger}_{val} (y) \beta \Psi_{m{^0}} (y)}
{E_{m^{0}} - E_{val}} \right .
\nonumber \\  & & \hspace{3cm} \;+\;\left . \frac{1}{2}
\Psi^{\dagger}_{n} (x) \Psi_{m^{0}} (x)
\Psi^{\dagger}_{m^{0}} (y) \beta \Psi_{n} (y)
{\cal R}_{\cal M} (E_n, E_m^{0}) \right ], \nonumber \\
 {\cal N}_1 (t) & = & \frac{N_c}{6} \sum_{n,m}
\int \; d^3x j_0 (Qr) \int d^3y \left[
\frac{\Psi^{\dagger}_{n} (x) \beta\vec{\tau} \Psi_{val} (x) \cdot
\Psi^{\dagger}_{val} (y) \beta \vec{\tau} \Psi_{n} (y)}
{E_n - E_{val}} \right .
\nonumber \\  & & \hspace{3cm} \;+\; \left . \frac{1}{2}
\Psi^{\dagger}_{n} (x) \beta \vec{\tau} \Psi_{m} (x) \cdot
\Psi^{\dagger}_{m} (y) \beta \vec{\tau} \Psi_{n} (y)
{\cal R}_{\beta} (E_n, E_m)
\right ],
\nonumber \\
{\cal N}_2 (t) & = & \frac{N_c}{6} \sum_{n, m^{0}}
\int d^3 x \;j_0 (Qr) \int d^3 y
\left [\frac{\Psi^{\dagger}_{m^{0}} (x)\beta \Psi_{val} (x)
\Psi^{\dagger}_{val} (y) \beta \Psi_{m{^0}} (y)}
{E_{m^{0}} - E_{val}} \right .
\nonumber \\  & & \hspace{3cm} \;+\;\left . \frac{1}{2}
\Psi^{\dagger}_{n} (x) \beta \Psi_{m^{0}} (x)
\Psi^{\dagger}_{m^{0}} (y) \beta \Psi_{n} (y)
{\cal R}_{\beta} (E_n, E_m^{0}) \right ], \nonumber \\
{\cal N}_0 (t) &=& \frac{3N_c}{2}
\sum_{n,m}
\int  d^3x j_0 (Qr) \int d^3y \left[
\frac{\Psi^{\dagger}_{n} (x) \beta\Psi_{val} (x)
\Psi^{\dagger}_{val} (y) \beta \Psi_{n} (y)}
{E_n - E_{val}} \right .
\nonumber \\  & & \hspace{3cm} \;+\; \left . \frac{1}{2}
\Psi^{\dagger}_{n} (x) \beta \Psi_{m} (x) \cdot
\Psi^{\dagger}_{m} (y) \beta \Psi_{n} (y)
{\cal R}_{\beta} (E_n, E_m)
\right ].
\end{eqnarray}
The subscripts $i$ and $p$ in the collective part are
$i=1,2,3$ and $p=4,5,6,7$, respectively.
$I_i$ and $K_i$ are respectively the moments of inertia and
anomalous moments of inertia~\cite{betal}.  When $t\rightarrow 0$,
${\cal K}_i (t) $ become $K_i$.  The $\Sigma_{SU(2)}$ corresponds to
the $\pi N$ sigma term in SU(2)~\cite{betal} at $t=0$,
which can be obtained by the Feynman-Hellman theorem
\begin{equation}
\Sigma_{SU(2)} = \left. \bar{m}
\frac{\partial E(\bar{m})}{\partial \bar{m}}\right|_{\bar{m}=0},
\end{equation}
where $E$ stands for the classical soliton energy.
The regularization functions $R (E_n)$, $R_{\cal M} (E_n, E_m)$,
$R_{\beta} (E_n, E_m)$
\footnote{$R_{\cal M} (E_n,E_m)$ is not actually a
regularization function, since ${\cal K}_i$ come from the imaginary
part of the action.  It does not depend on the cut-off parameter.}
are defined  by
\begin{eqnarray}
{\cal R} (E_n)  & = & \int \frac{du}{\sqrt{\pi u}}
\phi (u;\Lambda_i) |E_n| e^{-uE^{2}_{n}},
\nonumber \\
{\cal R}_{\cal M} (E_n, E_m) & = &
\frac{1}{2}  \frac{ {\rm sign} (E_n)
- {\rm sign} (E_m)}{E_n - E_m},
\nonumber \\
{\cal R}_{\beta} (E_n, E_m) & = &
\int^{\infty}_{0} \frac{du}{2\sqrt{\pi u}} \phi (u;\Lambda_i)
\frac{E_n e^{-u E^{2}_{n}} - E_m e^{-u E^{2}_{m}}}{E_n - E_m},
\end{eqnarray}
respectively.
The $\langle\rangle_N$ stands for the expectation value of the
Wigner $D$ functions in collective space apanned by $A$.
The expectation values of the $D$ functions can be evaluated by
SU(3) Clebsch-Gordan coefficients found in Refs.~\cite{Swart,Mcnamee}.
With SU(3) symmetry explicitly broken by $m_s$, the collective part is no
longer SU(3)-symmetric.  Therefore, the eigenstates of the
hamiltonian are not in a pure octet or decuplet but mixed states.
Since we treat the strange quark mass $m_s$ perturbatively,
we can obtain the mixed SU(3) baryonic states as follows:
\begin{equation}
| 8, N \rangle \;=\; | 8, N \rangle \;+ \;
c^{N}_{\bar{10}} | \bar{10}, N \rangle
\;+\;c^{N}_{27} | 27, N \rangle
\end{equation}
with
\begin{equation}
c^{N}_{\bar{10}} \;=\; \frac{\sqrt{5}}{15}(\bar{\sigma} - r_1)
I_2 m_s,
c^{N}_{27} \;=\; \frac{\sqrt{6}}{75}(3\bar{\sigma} + r_1 - 4r_2)
I_2 m_s.
\label{Eq:g2}
\end{equation}
The constant $\bar{\sigma}$ is related to the $\Sigma_{SU(2)}$
by $\Sigma_{SU(2)} = 2/3 (m_u + m_d)\bar{\sigma}$.  $r_i$ denotes
the ratio $K_i / I_i$.

Since the Cheng-Dashen point is out of the physical region,
it is necessary to extrapolate to the region $t>0$.
This can be done by the analytic continuation of the $|\vec{q}|$,
{\em i.e.} $|\vec{q}| \rightarrow i|\vec{q}|$ so that
we may have the positive $t$ up to the Cheng-Dashen point
($t=2m^{2}_{\pi}$).  The analytic continuation
above the threshold $t=4m^{2}_{\pi}$ is not valid in our model,
since above this threshold, the correlation between mesonic
clouds is getting important~\cite{phs}.
Hence, in this work, we only evaluate the scalar form factor
from the Cheng-Dashen point to the physical channel
(space-like region: $t<0$).
\section{Numerical Results and Dicussion}
In order to calculate the $\sigma_{\pi N} (t)$
numerically, we take advantage of
the Kahana-Ripka discretized basis~\cite{kr}.  Figure 1 shows
the scalar form factor as a function of the constituent quark mass $M
=M_u = M_d$.  The $\sigma (t)$ decreases as the $M$ increases,
in particular, below $t=0$.  As a result, the difference between
the $\sigma (2m^{2}_{\pi})$ and $\sigma (0)$ changes drastically
when we increase the $M$ from $370\;\mbox{MeV}$ to $450\;\mbox{MeV}$, as
shown in Table 1.  We select the $M=420\;\mbox{MeV}$
for the best fit as we did for other observables.
The error bar presented in Fig. 1 stands for the empirical
analysis due to Gasser, Leutwyler,
and Sainio~\cite{gls}, {\em i.e.} $\sigma (0) =45\pm 8 \; \mbox{MeV}$.
Our numerical prediction is in a remarkable agreement with
Ref.~\cite{gls}.  It is also interesting to see how the $m_s$ corrections
contribute to the scalar form factor.  As shown in Fig. 2, the $m_s$
corrections are very small.  At $t=0$, the $m_s$ corrections contribute
to the $\sigma$ term about $2\%$ which is negligible.
However, the $m_s$ corrections
play a significant role of reducing remarkably the large
strangeness contribution $\langle N | \bar{s} s | N \rangle$
arising from the leading term and rotational $1/N_c$
corrections.  With the $m_s$ corrections taken into account,
we obtain $y=0.27$ in case of the $M=420\mbox{MeV}$, which
agrees with the empirical value $y\simeq 0.2$~\cite{gls} within
about $30\%$, whereas we
have $y=0.48$ without the $m_s$ corrections.  It is already known that
the explicit symmetry breaking term quenchs the
$\langle N | \bar{s} s | N \rangle$~\cite{bjm,hatsuda,kk}.

The difference $\Delta \sigma = \sigma(2m^{2}_{\pi})
-\sigma(0)$ we have obtained is $18.18\mbox{MeV}$.  This value is
very close to what Gasser and Leutwyler extracted~\cite{gl},
$\Delta \sigma = 15.2\pm 0.4 \mbox{MeV}$.  The tangent of
the scalar form factor at $t=0$ is known to be related to
the scalar square radius.  It is almost
two times larger than the electric one, {\em i.e.}
the $\langle r^2\rangle^{S}_{N} \simeq 1.6 \mbox{fm}^2$
while $\langle r^2\rangle^{E}_{N} \simeq 0.74\mbox{fm}^2$.
The prediction of our model for the $\langle r^2\rangle^{S}_{N}$
is $1.5 \mbox{fm}^2$ which is almost the same as
obtained by Gasser and Leutwyler.
It implies that the tail of the
scalar density is of great importance.  In Fig. 3 we can find a
long-stretched and strong tail in the sea contribution to the scalar
density.  This tail is due to the mesonic clouds arising from the
Dirac sea polarization.
Moreover, the sea contribution in the scalar density is
large, compared with the other densities such as
electromagnetic densities~\cite{wkg,wakamatsu,cggp}.

The other interesting quantities are presented in table 1.
$\sigma_0$ is the condensate of the singlet scalar quark operator
in the nucleon:$\sigma_0 = \bar{m} \langle N |
\bar{u}u+\bar{d}d + \bar{s}s| N \rangle$
$R_s$ is defined by
$R_{s} = \langle N | \bar{s}s | N \rangle /
 \langle N | \bar{u}u+\bar{d}d + \bar{s}s| N \rangle$.
\section{Summary and Conclusion}
We have discussed the scalar form factor with related
quantities in the SU(3) NJL soliton model.  The results we have obtained
are in a good agreement with empirical data~\cite{gls,gl}.
The reliable strangeness contents of the nucleon in the scalar
channel is obtained by taking into account the $m_s$ corrections,
since they suppress the excess of $\langle N|\bar{s}s | N \rangle$
due to the leading order and rotational $1/N_c$ contributions.
In contrast to Refs.~\cite{bft,bass} suggesting no strangeness
contribution, our model favors $y=0.27$.
The large value of the $\langle r^2\rangle^{S}_{N}$
is caused by the pronounced long ranging tail
which can be identified with the pion and kaon clouds.
\section*{Acknowledgement}
The authors would like to thank M. Polyakov for helpful discussions.
This work has partly been supported by the BMFT, the DFG,
the COSY--Project (J\" ulich) and Department of Energy
grant DE-FG02-88ER40388.  One of us (AB) would like to thank
the {\it Alexander von Humboldt   Foundation} for a Feodor Lynen grant.

\begin{table}
\caption{The physical quantities related to the scalar form factor.
The empirical data come from ~Ref.[3,17].}
\begin{tabular}{c|c|c|c|c|c|c|c}
$M$ & \multicolumn{2}{c|}{$370$ MeV} & \multicolumn{2}{c|}{$420$ MeV}
&  \multicolumn{2}{c|}{$450$ MeV} & Exp \\ \cline{1-7}
$m_s$ [MeV] & 0 & 156.75 & 0 & 148.49 & 0 & 145.35 & \phantom{} \\ \hline
$\sigma_{\pi N}$[MeV] & 43.09 & 44.71 & 40.01 & 40.80
& 38.22 & 38.69 & $45\pm 8$ \\
$\sigma_{0}$[MeV] & 53.25 & 49.25 & 49.58 & 46.24
& 47.37 & 44.35 & \phantom{} \\
$\sigma_{8}$[MeV] & 22.77 & 35.63 & 20.87 & 29.92
& 19.92 & 28.37 & $\phantom{}$  \\
$y$ & 0.47 & 0.20 & 0.48 & 0.27 & 0.48 & 0.29 & $0.2\pm 0.2$ \\
$R_{s}$ & 0.19 & 0.09 & 0.19 & 0.12 & 0.19 & 0.13 & \phantom{} \\
$\Delta\sigma$[MeV] &32.29&33.37&18.36&18.18&14.23&13.84&$15.2\pm0.4$  \\
$ \langle r^2\rangle^{S}_{N}$ &1.94&1.87&1.56&1.50&1.40&1.34&1.6
\end{tabular}
\end{table}

\vfill\eject
%====================================================================
\newpage
\pagebreak
\end{document}